\documentclass[]{spie}  

\usepackage{amsmath,amsfonts,amssymb}
\usepackage{graphicx}
\usepackage[colorlinks=true, allcolors=blue]{hyperref}

\title{Using raytracing to derive the expected performance of STELLA's SES-VIS spectrograph}
\author[a]{Michael Weber}
\author[a]{Klaus G.~Strassmeier}
\author[a]{Manfred Woche}
\author[a]{Ilya Ilyin}
\author[a]{Arto J\"arvinen}
\affil[a]{Leibniz-Institut for Astrophysics Potsdam (AIP), An der Sternwarte 16, D--14482 Potsdam, Germany}

\authorinfo{Send correspondence to M. Weber: E-mail: mweber@aip.de}

\newcommand{\ms}{m\,s$^{-1}$}

\begin{document} 
\maketitle

\begin{abstract}
The visual STELLA \'echelle spectrograph (SES-VIS) is a new instrument for the STELLA-II telescope at the Iza\~na observatory on Tenerife. Together with the original SES spectrograph - which will still be used in the near IR - and a new H\&K-optimized spectrograph, which is currently in the design phase, it will extend the capabilities of STELLA with the follow up of planetary candidates from space missions (TESS, PLATO2). SES-VIS is optimized for precise radial velocity determinations and long term stability.
We have developed a ZEMAX based software package to create simulated spectra, which are then extracted using our new data reduction package developed for the PEPSI spectrograph. 
The focus in this paper has been put on calibration spectra, and the full range of available calibration sources (flat field, Th-Ar, and Fabry-P\'erot etalon), which can be compared to actual commissioning data once they are available.
Furthermore we tested for the effect of changes of the environmental parameters to the wavelength calibration precision.
\end{abstract}


\section{INTRODUCTION}
\label{sec:intro}  

To measure precise radial velocities with a spectrograph essentially means to measure the position of the edges of bright and dark features on the CCD. This is most obvious for calibration spectra, which are either line-spectra from elements like Thorium, or artificial line spectra from a laser comb or a stabilized Fabry-P\'erot etalon (FPE). These edges can be determined approximately (depending on contrast) with an accuracy of 10\% of the width of a pixel, which is in case of a spectrograph with a resolving power of 5$\times 10^4$ approximately 100\,\ms. To reach the goal of \ms\ accuracy, one needs many lines, in this case about 1$\times 10^4$. For calibration light sources, if light sources like ThAr are not sufficient any more, laser combs or stabilized Fabry-P\'erot etalons that can be matched to the spectrograph parameters are the solution. For science observations, the only way to increase the number of lines is the increase of the wavelength range, and to some extent for slowly rotating stars the increase of spectral resolution. The actual factor defining the radial velocity precision is the slope of the spectral lines.

Hatzes \& Cochran\cite{HatzesCochran1992} used this fact to derive a formula for the expected radial velocity precision in terms of spectral resolution, signal/noise ratio and wavelength coverage:
\begin{equation}
\label{formula:hatzes}
\sigma = 1.45 \times 10^9 (S/N)^{-1} R^{-1} B^{-1/2} (m/s)
\end{equation}
where $S/N$ is the signal-to-noise ratio of the spectrum, $R$ the spectral resolution $\lambda / \Delta \lambda$, and $B$ the spectral coverage in \AA.  For the present case of SES-VIS with the parameters listed in Sect.~\ref{sec:sesvis}, the expected precision from this formula would only depend on $S/N$ and would vary between 1\,m/s for a S/N of 600 and 6\,m/s for a S/N of 100. 

A more stringent formula for the expected radial velocity precision has been derived by Bouchy et al. (2001)\cite{Bouchyetal2001}, taking the spectral type and the broadening of the star into account in addition to S/N and spectral resolution. A good recent summary of the various approaches has been written by Figueira (2018)\cite{figueira2018}.

\section{FABRY-PEROT UNIT}

\begin{figure}[ht]
\includegraphics[width=0.66\textwidth]{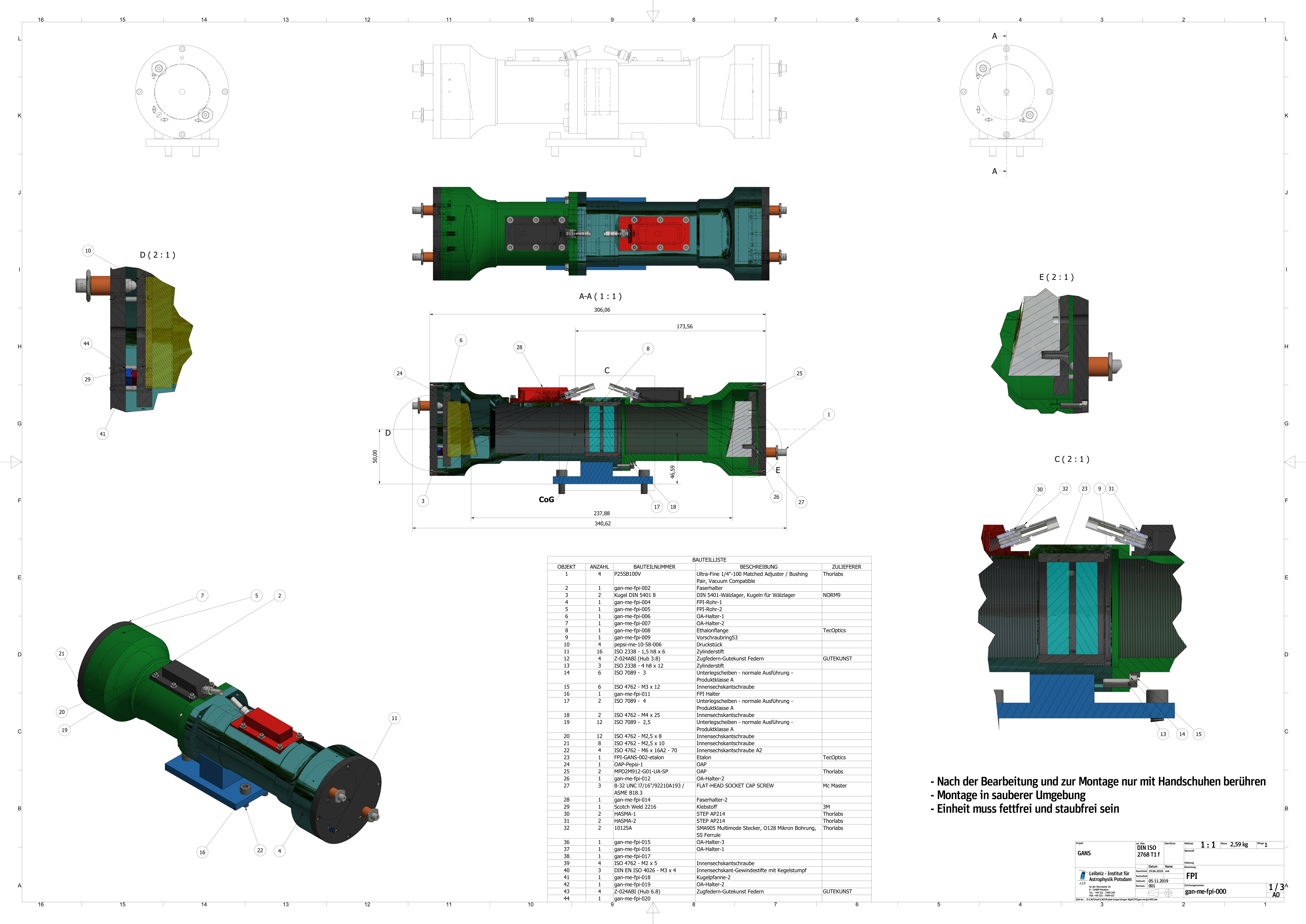}
\includegraphics[width=0.34\textwidth]{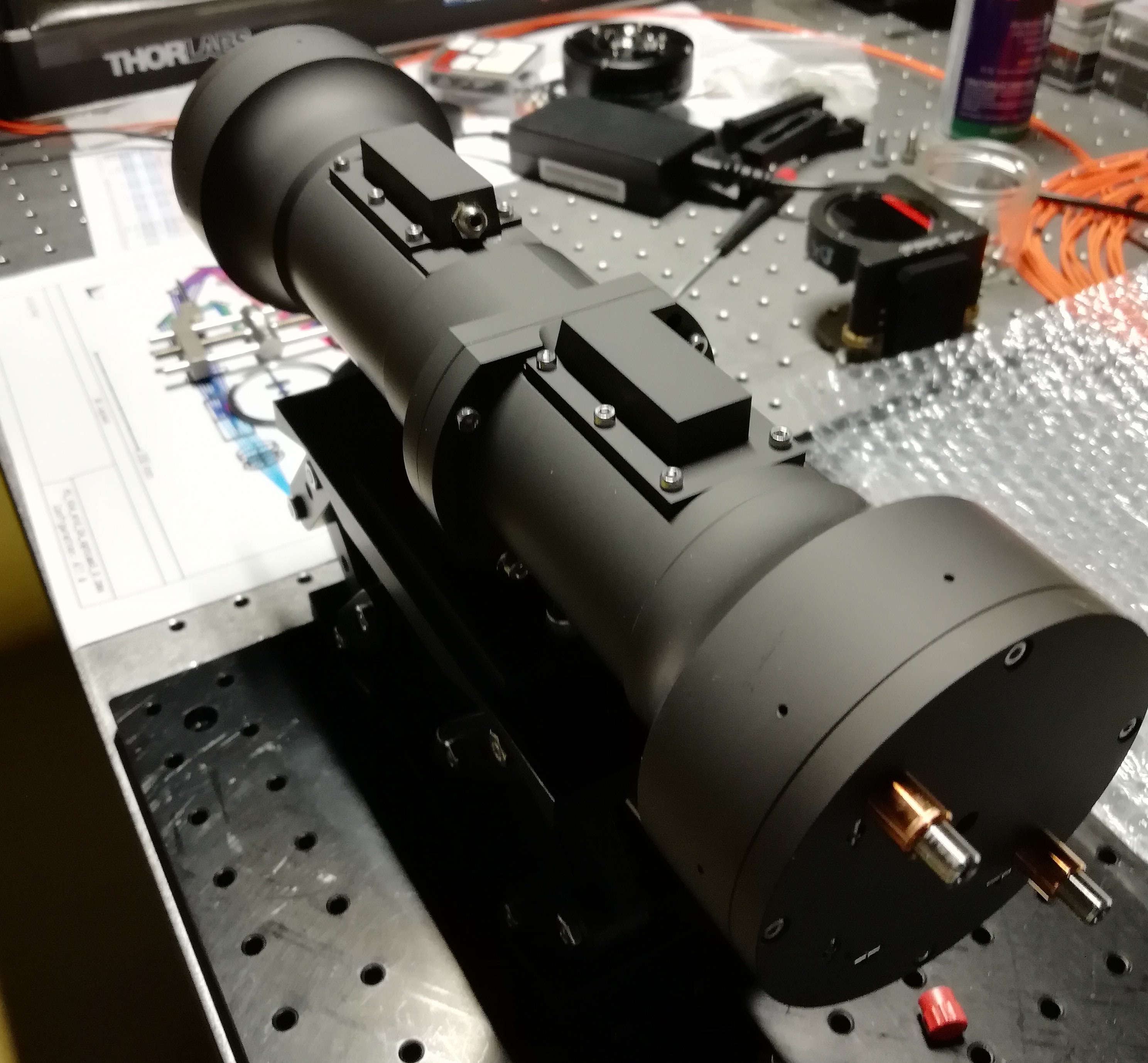}
\includegraphics[width=\textwidth]{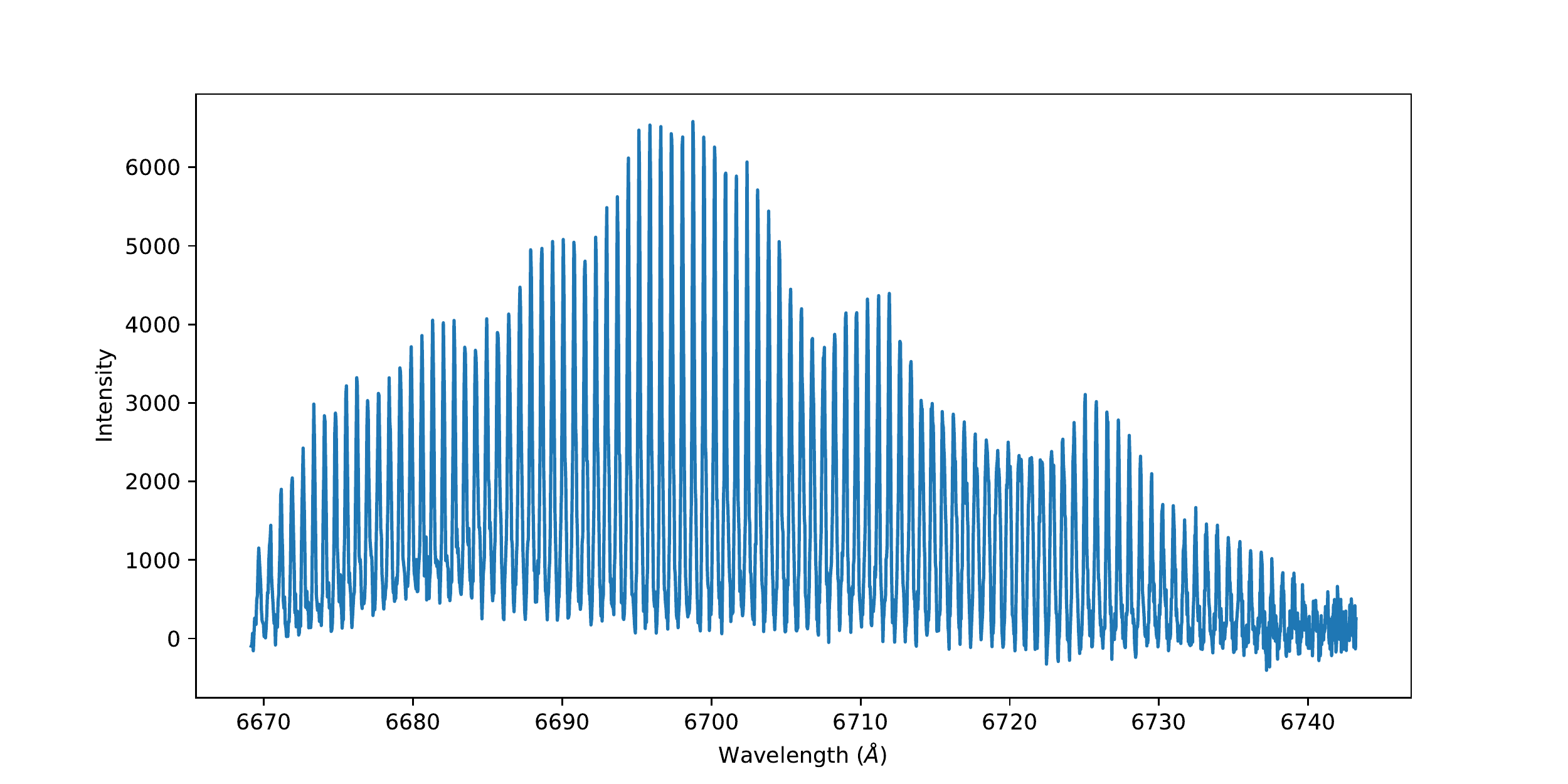}
\caption{\label{fig:fpi}
Design of the Fabry-P\'erot etalon (FPE) unit of SES-VIS. 
{\em Top:} Assembly drawing of the FPE unit at the left, an image of the assembled unit during lab testing on the right. Fiber input and output is located at the top, the etalon is located in the center of the unit.
{\em Bottom:} One order of the FPE-spectrum recorded with a badly aligned laboratory \'echelle spectrograph.
}
\end{figure}

Since a laser-comb calibration light source was out of reach when we started this project, we decided to build a calibration source using a Fabry-P\'erot etalon, similar to the one used in PEPSI. We used two off-axis parabolic mirrors (OAPs), one to collimate the f/3 output of an optical fiber fed with halogen light, the other to re-focus the beam onto a fiber which feeds the calibration light path of the spectrograph. For the first OAP we used a high-quality ($\lambda$/10) Zerodur mirror with broadband silver coating to make sure the etalon is well illuminated. The second OAP was manufactured out of solid aluminum to match the properties of the first OAP. The etalon has a cavity size of 3.08\,mm and a beam size of 25\,mm, resulting in an effective finesse (neglecting the aperture finesse of the feed fiber) of 23 to 25 at our wavelength range of 470\,nm to 690\,nm. 

The unit was built as a vacuum-compatible tube-structure (see Fig,~\ref{fig:fpi} top), which will be put inside the vacuum vessel of SES-VIS to keep the spectrum stable. A prototype was assembled at the AIP, aligned, and we used a small laboratory spectrograph to test the functionality of the unit (see Fig,~\ref{fig:fpi} bottom). Unfortunately, alignment problems of the laboratory spectrograph hindered us to make detailed analysis of the performance of the etalon. 

In total we will get about 175 FPE lines per order in all 42 orders of SES-VIS, resulting in about 7000 lines in total. Assuming 100\,\ms accuracy per line, the calibration accuracy will achieve about 1\,\ms, matching the best-case stellar radial velocity sensitivity derived above. In order to keep the wavelengths of the FPE-peaks constant to better than 1\,\ms, the pressure stability inside the FPE must be better than 0.01\,mbar.

\section{THE SES-VIS SPECTROGRAPH}
\label{sec:sesvis}  

\begin{figure}[ht]
\includegraphics[width=\textwidth]{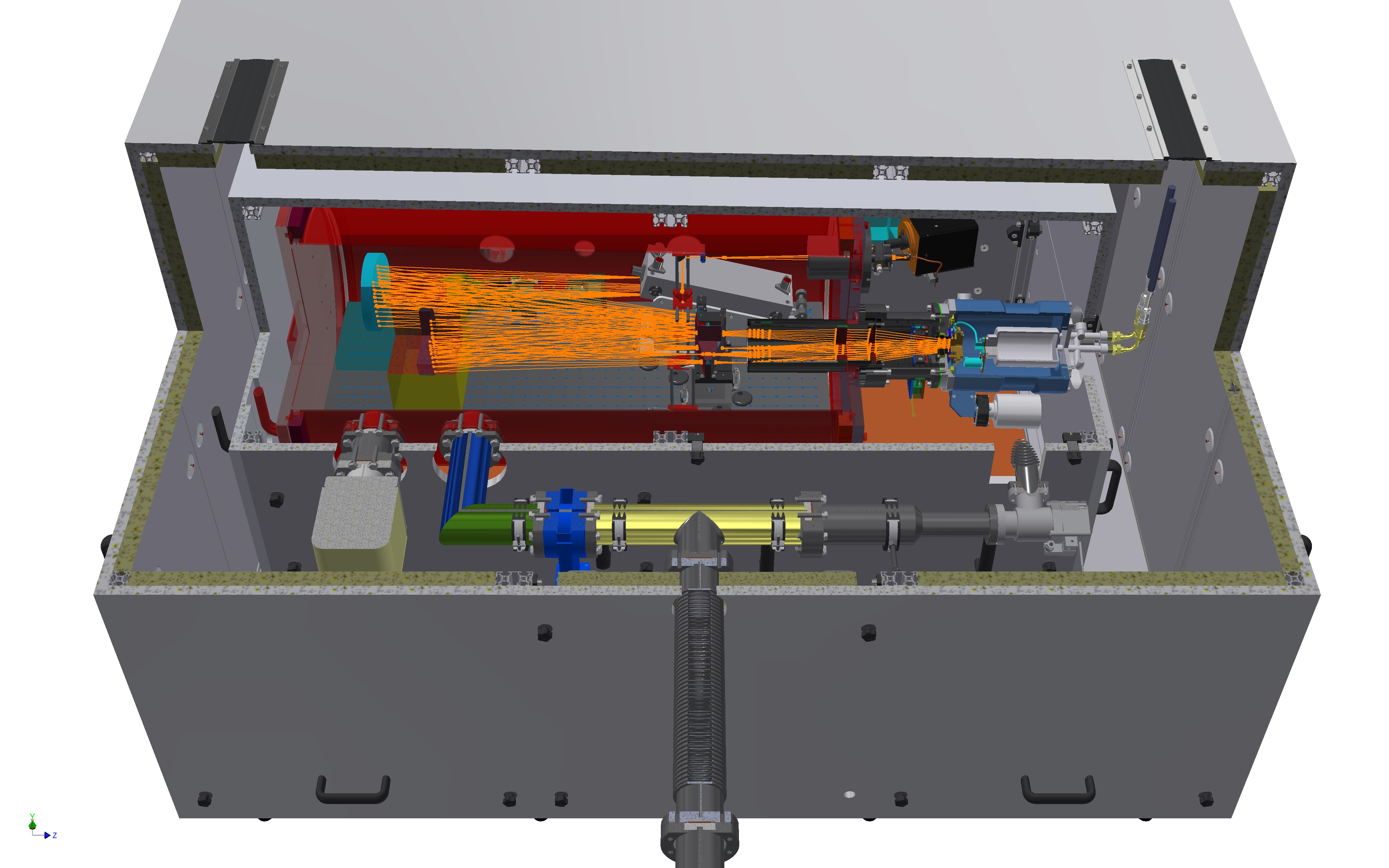}
\caption{\label{fig:sesvis}
System design of the SES-VIS spectrograph. The innermost part is the vacuum vessel with a diameter of 500\,mm, surrounded by two insulated enclosures. The volume between these two enclosures is actively kept at a constant temperature. The whole unit is located in a commercially air-conditioned room.
}
\end{figure}

The new SES-VIS spectrograph was designed for the solar telescope GREGOR and known as GANS\cite{2018SPIE10702E..6LJ}. It uses a classical white pupil optical design. 
Its design goals were to measure precise radial velocities, it is therefore contained inside a vacuum vessel for stability, which again is surrounded by an actively temperature-controlled enclosure (Fig.~\ref{fig:sesvis}). 
We use a R4-\'echelle grating, with orders 89 - 130 for wavelengths from 468\,nm to 691\,nm. In order to detect possible shifts between wavelength calibration exposures, a spectrum through a FPE is put alongside all calibration and science exposures (Fig.~\ref{fig:sesvis:simul:fp}). 
More details can be found in the presentation from SPIE 2018\cite{2018SPIE10705E..1EW}.

\section{SIMULATED SPECTRA}
\label{sec:simulation}  

\begin{figure}[t]
\includegraphics[width=\textwidth]{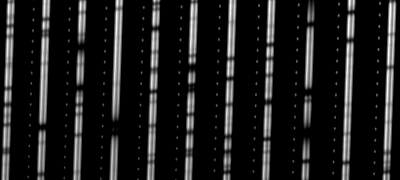}
\caption{\label{fig:sesvis:simul:fp}
Small excerpt of one two-dimensional spectrum from our simulations: Next to the science spectrum, which is double-humped due to the image slicer, is the spectrum of the FPE with its evenly distributed peaks.
}
\end{figure}

Using the optics design program Zemax we used the as-built optical design of SES-VIS to compute 2D-\'echelle spectra from standard one-dimensional spectra\cite{echmod,2016SPIE.9911E..2DG}. Recent versions of Zemax OpticStudio feature a well documented programming interface, which allows - among other things - to create standalone python programs that use OpticStudio to trace rays. The python program reads in the model spectrum and the order list of GANS specifying start and end wavelength for each spectral order that fits on the detector surface. For each spectral order, the flux is divided into small wavelength chunks (matching the sampling of the spectrograph), and each of these chunks is propagated through the spectrograph. The relative intensity is preserved by scaling the propagated rays with the incoming intensity. The rays that end up at the detector surface are assigned to their respective pixel, and the 2D-\'echelle image is obtained after going through the complete order list. We use a fairly high number of rays (10000) per sampling element (one pixel in dispersion direction), which are constructed by uniformly distributing them across the field (the optical fibre diameter) and pupil of the spectrograph.

Our primary spectrum for all further analysis was a model spectrum of the sun. We used the solar atmosphere model from MARCS\cite{2008A&A...486..951G}, a line list from VALD3\cite{2015PhyS...90e4005R}   and the Turbospectrum\cite{1998A&A...330.1109A} code to compute spectra the model spectrum of the sun. After the model spectrum was broadened to account for rotation and macroturbulence, it became our reference spectrum for further analysis.

To analyze the results we used our data reduction package SDS4PEPSI\cite{2018A&A...612A..44S}, extracting the spectral orders from for both the science spectrum as well as the FPE-spectrum. For the purpose of data reduction we constructed flat-field images from one model with a constant intensity, and wavelength reference images using a high-resolution PEPSI\cite{2018A&A...612A..44S} ThAr spectrum (see Fig.~\ref{fig:sesvis:simul:fp}). 

During the analysis of our data it became clear that we need to add features to our data reduction package. Currently it is not possible to supply a reference line list on the fly, as would be necessary to use the positions of the FPE-peaks to set the wavelengths of the FPE-spectrum. We overcame this limitations by using a ThAr-Spectrum in place of the FPE reference spectrum. We will also measure this during commissioning of the spectrograph, but during routine operations only light from the FPE-unit can be placed simultaneously onto the detector. It is also not possible currently to use optimal extraction\cite{Horne_1986}, since this fails on the FPE spectrum but can not be turned off selectively for individual slices, and our simultaneous calibration spectrum is treated as an extra slice. 
Solutions for these problems are under way and will be ready for lab-commissioning. 

\section{RESULTS}

\begin{figure}[t]
\includegraphics[width=\textwidth]{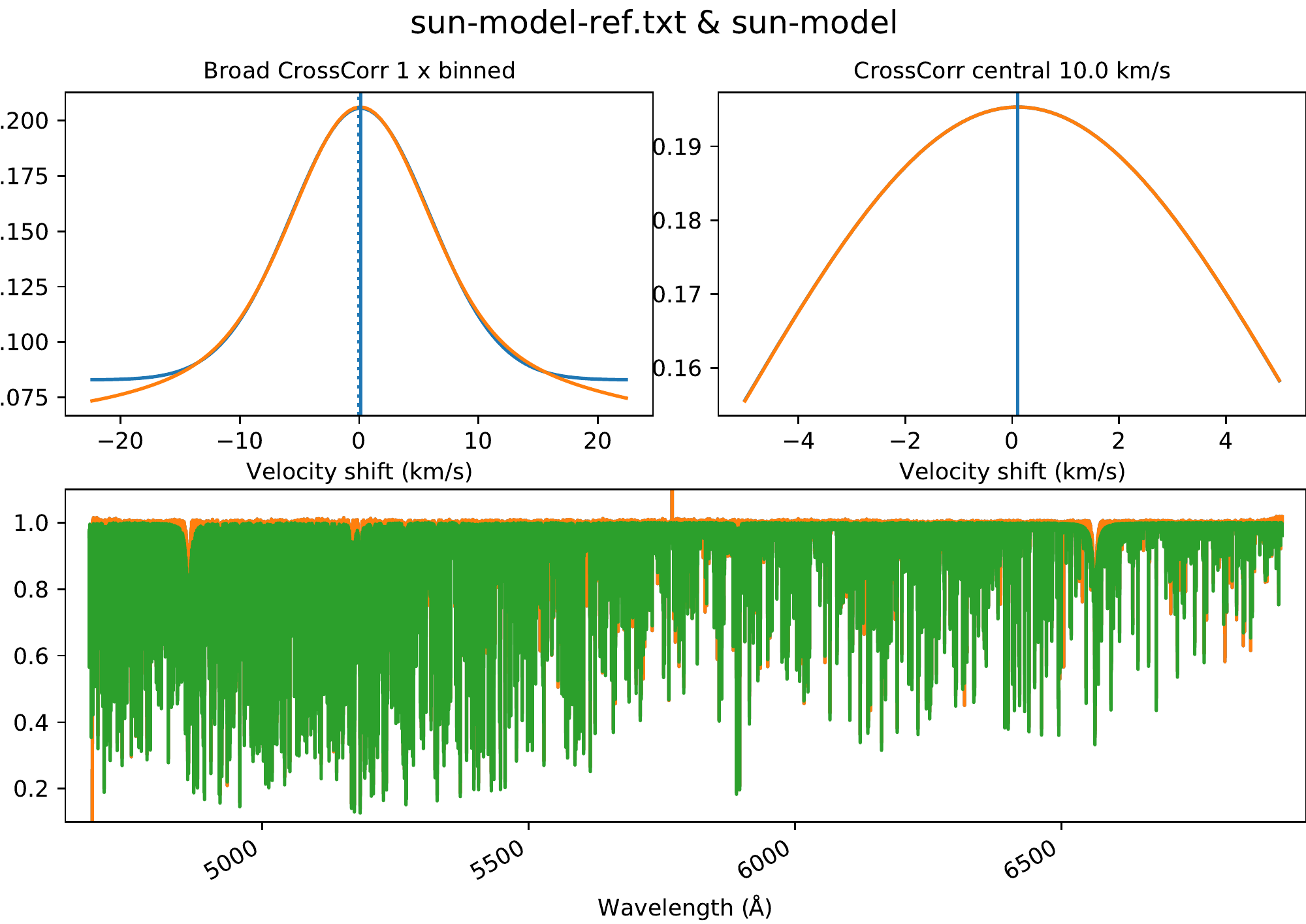}
\caption{\label{fig:crosscorr}
Example graph from the cross-correlation procedure: The science spectrum is first cross-correlated with a template using a standard routine, re-binning all points to the same velocity scale (top left). In a second step a cross-correlation is performed changing only the sampling of the template (top right).
The bottom panel shows the science spectrum and the shifted and spun up template spectrum.
}
\end{figure}

We created a set of simulated spectra to test the data analysis software, and to see up to what quality of the vacuum the radial velocity can be derived at a nominal precision of about 1\,\ms, which is roughly our design goal since at that resolution and wavelength range it is the best precision one can obtain from a well-behaved stellar spectrum (see Sect.~\ref{sec:intro} above). Combinations are written as ThAR/FPE, meaning that ThAr-light was used for the science orders, and FPE-light was used for the calibration orders. This set consisted of flat field spectra in all combinations (Flat/zero, zero/Flat, Flat/Flat), ThAr-calibration spectra (ThAr/ThAr and Thar/FPE), and s series of Sun/FPE spectra at pressure settings inside the vacuum vessel between 0.01 and 100\,mbar. 

We plan to run the vessel at a pressure of 10$^{-4}$\,mbar, but want to know at what level we would need to start pumping in order to keep up our wavelength stability. The FPE is located in the same vacuum vessel, and a calibration accuracy of 1\,\ms\ demands a pressure stability of better than 0.01\,mbar, otherwise changes in the FPE-peak wavelengths would dictate the precision of the simultaneous calibration.

All spectra are extracted using standard extraction, the science and calibration spectra are treated as two slices and are extracted together, and the wavelength solution is obtained using both spectra simultaneously also. The spectra are then separated out into two sets of spectra, science spectra and calibration spectra, normalized using per-order spline-fits, and merged into single one-dimensional spectra. For simplicity we did not make a global fit to set the continuum, nor did we iterate the continuum to match the order-overlaps, thus small continuum-errors are introduced during this procedure.

\begin{figure}[t]
\includegraphics[width=\textwidth]{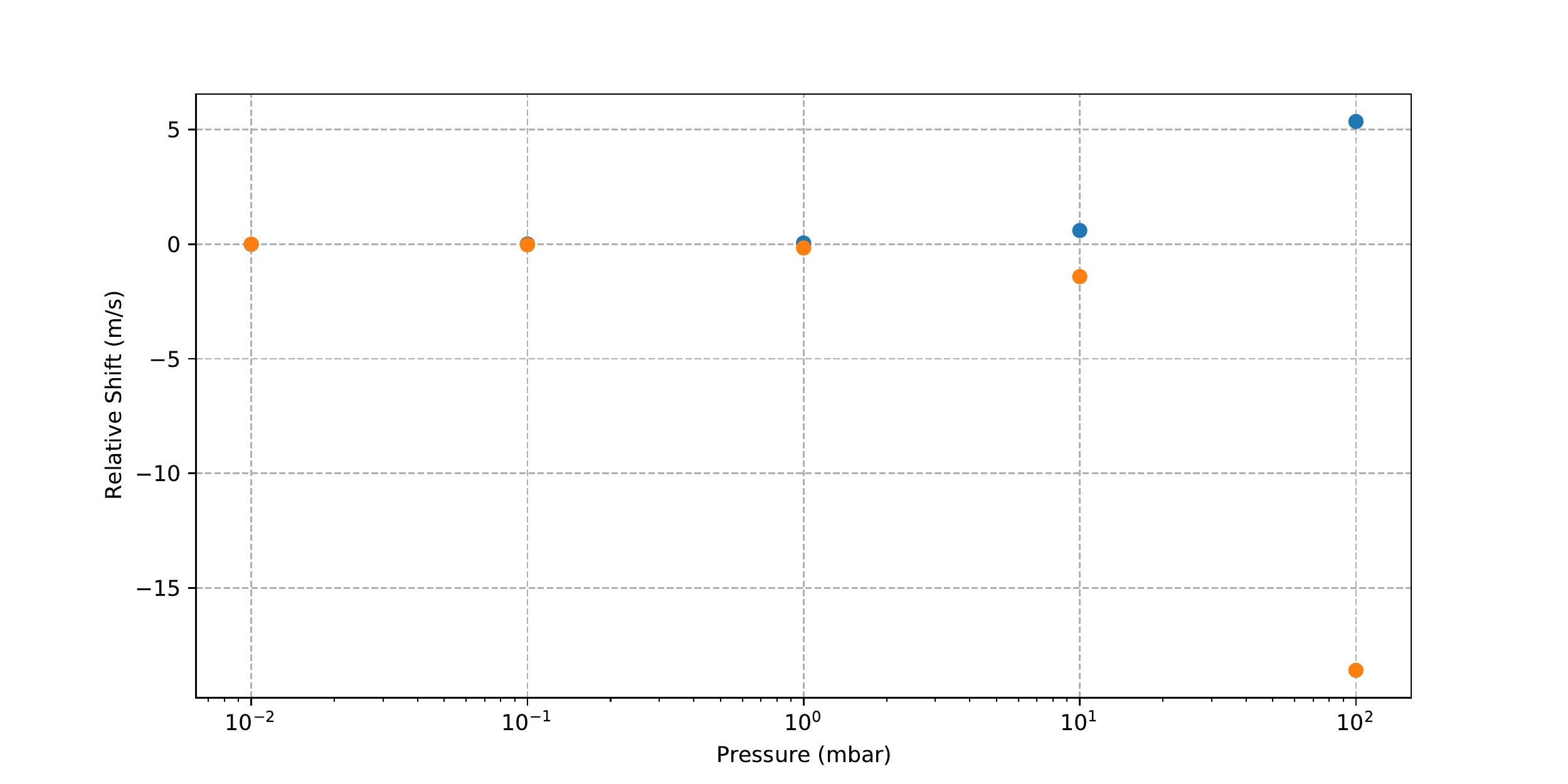}
\caption{\label{fig:crosscorr:comp}
Result from cross-correlating the extracted solar spectrum and the FPE spectrum with its corresponding model, at various pressure settings. A relative drift sets in at about 1\,mbar.
}
\end{figure}

As our data reduction software is not yet capable of directly measuring the stability of the FPE-peaks' positions, we independently derived the global shift of the solar spectrum using the extracted science spectra, and of the FPE-spectrum using the extracted calibration spectra. We used the same spectra as cross-correlation templates as we used as input for ray tracing. No additional noise was added to the images, the number of rays traced corresponds to a S/N of about 100 per pixel. 

Cross-correlation was carried out by a multi-purpose python script developed for merged STELLA and PEPSI spectra. The template can optionally be spun up to an arbitrary rotational velocity, and macro-turbulence and instrumental broadening can be folded in. Wavelength-regions of interest can be defined that leave out regions which are heavily affected by telluric lines, or to leave out (or focus on) very strong lines like Balmer-lines or Na-D, but here the whole spectral region was used. In a first step both template and science spectrum are re-sampled to a constant velocity step size and a regular correlation routine is used to obtain a first approximation of the peak position (top left of Fig.~\ref{fig:crosscorr}). In a second step, the natural sampling of the science spectrum is preserved, and the template spectrum is evaluated at these sampling wavelengths.  We then repeat this procedure for 5000 different velocity shifts, offset each by 1\,\ms centered around the peak derived in the first step, and derive the final offset by a fit to this function (top right of Fig.~\ref{fig:crosscorr}). 
In the bottom of Fig.~\ref{fig:crosscorr} the science spectrum is shown along with the shifted and broadened template spectrum. Naturally, since both are derived for the identical model spectrum, they are very similar, but a small waviness in the continuum setting is visible as well as systematic differences around strong lines.

Now, in the end, we finally get to compare the offsets derived from both the science spectra as well as the calibration spectra with respect to the vacuum-vessel's pressure (Fig.~\ref{fig:crosscorr:comp}). The offsets stay constant up to 0.1\,mbar, and only an insignificant difference is visible at 1\,mbar. From here onwards, the differences are much bigger than the desired accuracy of the spectrograph, as was expected.

This experiment shows that we are on the safe side with our planned pressure level at 10$^{-4}$\,mbar. The spectrograph itself tolerates two orders of magnitudes more than the FPE, which leaves some room for effects like residual temperature variations which are not covered here since we can not simulate them during lab-commissioning due to the lack of space for the thermal enclosure. And the extra margin of two orders of magnitude from our planned vacuum level of 10$^{-4}$\,mbar to a tolerable 0.01\,mbar suggests that we can refrain from pumping during the nights.

\acknowledgments 
The STELLA facility is funded by the Science and Culture Ministry of the German State of Brandenburg (MWFK) and the German Federal Ministry for Education and Research (BMBF), and is operated by the AIP jointly with the IAC.

\bibliography{aa_mnem,gans_verification} 
\bibliographystyle{spiebib} 

\end{document}